\documentstyle[twoside,fleqn,npb,epsfig]{article}
\newcommand{\eg}{{\it e.g.\ }}

\def\lsim{\mathrel{\rlap{\lower4pt\hbox{\hskip1pt$\sim$}}
    \raise1pt\hbox{$<$}}}         
\def\gsim{\mathrel{\rlap{\lower4pt\hbox{\hskip1pt$\sim$}}
    \raise1pt\hbox{$>$}}}         

\title{Rapidity Gaps from Colour String Topologies\thanks{Contribution to
the DIS 99 workshop proceedings}} 

\author{G.\ Ingelman$^{ab}$\thanks{~ingelman@desy.de},  
        A.\ Edin$^b$, R.\ Enberg$^a$, J.\ Rathsman$^c$, N.\ Timneanu$^a$\\ ~~ \\
$^a$~High Energy Physics, Uppsala University, Box 535, S-751 21 Uppsala, Sweden\\
$^b$~DESY, Notkestrasse 85, D-22603 Hamburg, Germany\\
$^c$~SLAC, Stanford, California 94309, USA\\
}
\begin{document}

\begin{abstract}
Diffractive deep inelastic scattering at HERA and diffractive $W$ and jet 
production at the Tevatron are well described by soft colour exchange models.
Their essence is the variation of colour string-field topologies giving both 
gap and no-gap events, with a smooth transition and thereby a unified 
description of all final states.   
\end{abstract}

\maketitle

The hard scale in diffractive hard scattering \cite{IS,StCroix} has provided 
the possibility to analyse rapidity gap events based on underlying parton 
processes
calculable in perturbation theory. Although this has been quite successful, 
perturbative QCD (PQCD) cannot give the complete solution since the rapidity 
gap connects to the soft part of the event where non-perturbative effects 
on a long space-time scale are important. 

In order to understand these non-perturbative effects and provide a unified 
description of all final states, we have developed models for the soft 
dynamics. These models are 
added to Monte Carlo generators ({\sc Lepto} \cite{Lepto} for $ep$ and 
{\sc Pythia} \cite{Pythia} for $p\bar{p}$), 
such that an experimental approach can be taken to classify 
events depending on the characteristics of the final state:
\eg gaps or no-gaps, leading protons or neutrons {\it etc}. 

\begin{figure}[tbh]
\begin{center}
\epsfig{width=1.0 \columnwidth,file=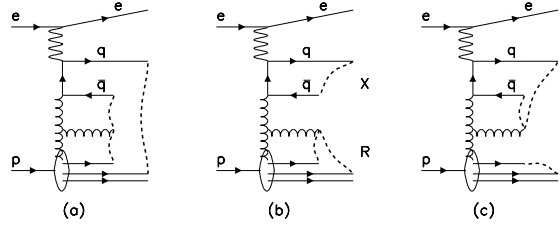} 
\end{center}
\vspace*{-10mm}
\caption[junk]{Gluon-induced DIS event with examples of colour string 
connection (dashed lines) of partons in 
(a) conventional Lund model based on the colour order in PQCD, and 
(b,c) after colour rearrangement through SCI or GAL mechanisms.
\label{fig:DIS}}
\vspace*{-5mm}
\end{figure}

The basic assumption of the models is that variations in the topology of the
confining colour  force fields (strings) lead to different hadronic final
states after  hadronisation (Fig.~\ref{fig:DIS}).
The PQCD interaction gives
a set of partons with a specific colour order. However, this order may change 
due to soft, non-perturbative interactions. 

\begin{figure}[th]
\center{\epsfig{width=0.75 \columnwidth,file=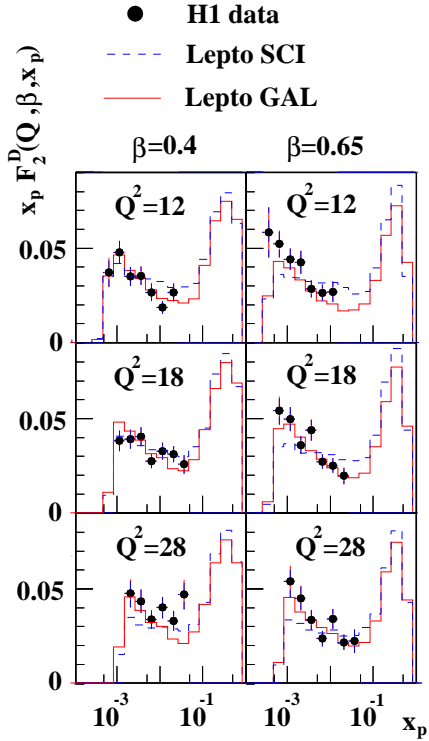}}
\vspace*{-8mm}
\caption{The diffractive structure function \cite{GAL-HERA}. 
}
\vspace*{-5mm}
\label{fig:f2d3}
\end{figure}

In the {\em soft colour interaction} (SCI) model \cite{SCI} it is assumed that
colour-anticolour, corresponding to non-perturbative gluons, can be  exchanged
between partons and remnants emerging from a hard scattering.  This can be
viewed as the partons interacting softly with the colour medium  of the proton
as they propagate through it,  which should be a natural part of the process in
which `bare' perturbative  partons are `dressed' into non-perturbative ones and
the confining colour  flux tube between them is formed.  The hard parton level
interactions are given by standard perturbative matrix  elements and parton
showers, which are not altered by softer non-perturbative effects. The unknown
probability to exchange a soft gluon between  parton pairs is
given by a phenomenological parameter $R$, which is the only free parameter of
the model.  With $R=0.5$ one obtains the correct rate of rapidity gap events
observed at  HERA and a quite decent description of the measured diffractive
structure  function \cite{H1} (Fig.~\ref{fig:f2d3}).

Leading neutrons are also obtained in agreement with experimental
measurements \cite{leading-pn}. In the Regge approach pomeron exchange would 
be used for diffraction, pion exchange added to get leading neutrons and still
other exchanges should be added for completeness. 
The SCI model provides a simpler description.

\begin{figure}[tb]
\center{
\epsfig{width=0.48 \columnwidth,file=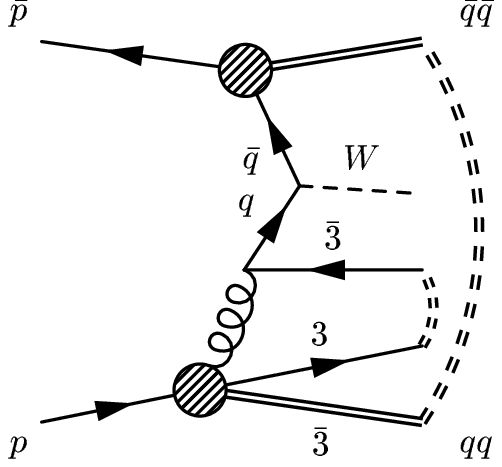}
\epsfig{width=0.48 \columnwidth,file=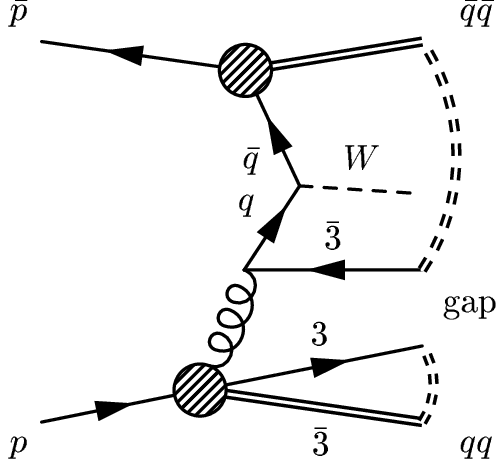}
\vspace*{-6mm}\\
\epsfig{width=0.9  \columnwidth,file=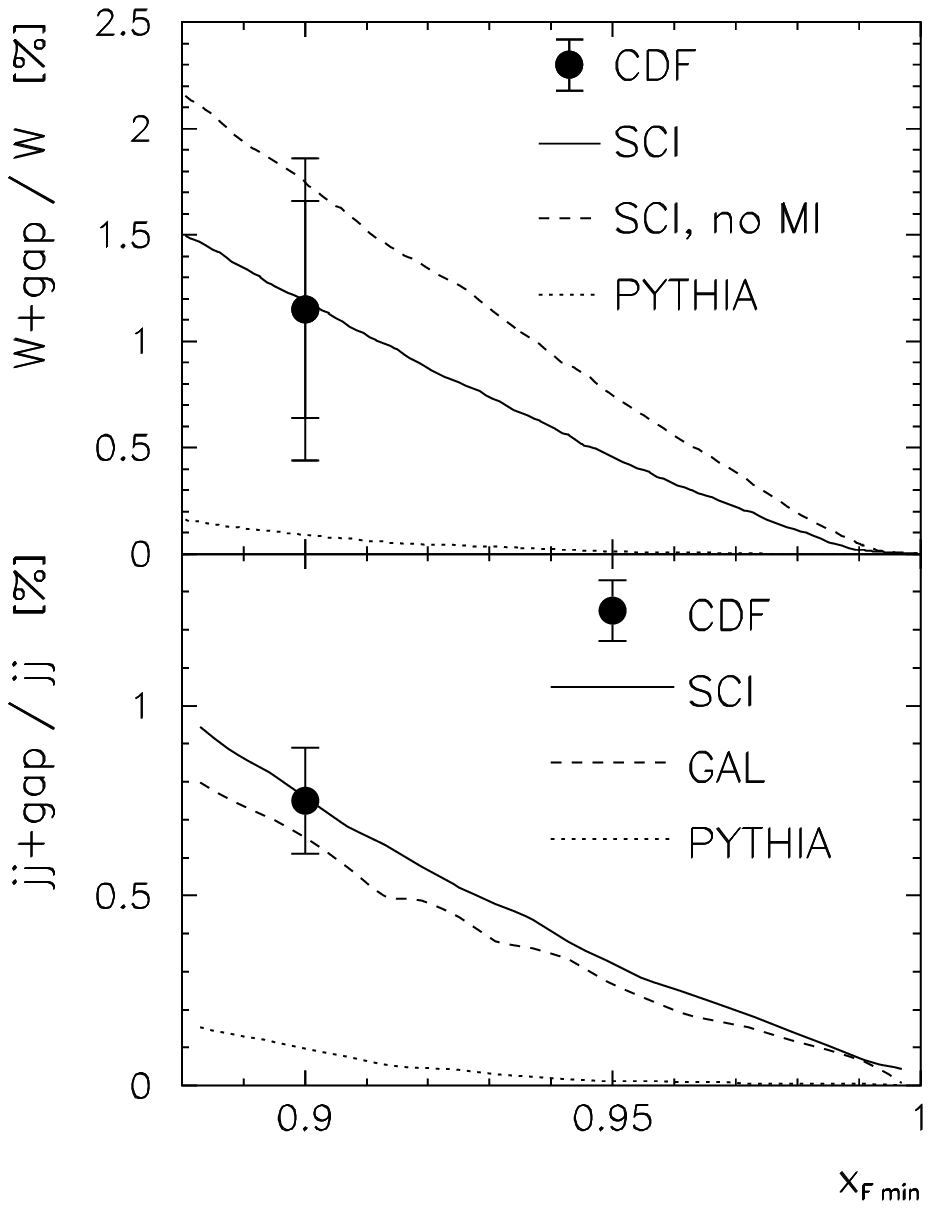}
}
\vspace*{-12mm}
\caption{$W$ production in $p\bar{p}$ with string topology before and after
colour rearrangement resulting in a gap. 
Relative rate of $W$ and di-jet events with a rapidity gap 
corresponding to diffraction with a leading proton with minimum $x_F=0.9$
in Tevatron data \cite{CDF-W+dijets}
and in the SCI and GAL models. Sensitivity to multiparton 
interactions (MI) and results without SCI or GAL is shown \cite{SCI-W-jets}.
}
\label{fig:Wsci}
\vspace*{-5mm}
\end{figure}

Applying the same SCI model to hard $p\bar{p}$ collisions 
one obtains production of $W$ and di-jets in association with rapidity gaps 
(Fig.~\ref{fig:Wsci}). Keeping the $R$-value obtained from 
gaps at HERA, the observed rates of diffractive $W$ and diffractive di-jet 
production at the Tevatron are reproduced (Fig.~\ref{fig:Wsci}). 
This is in contrast to the Pomeron model which, when tuned to HERA gap events, 
gives a factor $\sim 6$ too large rate at the Tevatron \cite{StCroix}. 

SCI does not only lead to rapidity gaps, but also to other striking effects. 
It reproduces \cite{SCI-psi} the observed rate of high-$p_{\perp}$ charmonium 
and bottomonium at the Tevatron, which are factors of 10 larger than 
predictions based on conventional PQCD. This is accomplished by the 
change of the colour charge of a $Q\bar{Q}$ pair (\eg from a gluon)
from octet to singlet. A quarkonium state can then be formed  using a simple 
model for the division of the cross-section below the threshold for open heavy 
flavour production onto different quarkonium states. 

An alternative to SCI is 
the newly developed {\em generalised area law} (GAL) model \cite{GAL} which,
based on a generalisation of the area law suppression $e^{-bA}$ with 
$A$ the area swept out by the string in energy-momentum space, 
gives modified colour string topologies through string reinteractions. 
The probability $P=R_0[1-exp(-b\Delta A)]$ 
for two strings pieces to interact depends on the area difference
$\Delta A$ which is gained by the string rearrangement. This
favours making `shorter' strings, \eg with gaps, whereas making `longer',
`zig-zag' shaped strings is suppressed. The
fixed probability $R$ in SCI is thus  replaced by a dynamical one, where the
parameter $R_0=0.1$ is chosen to reproduce the HERA gap event rate in a 
simultaneous fit to data from $e^+e^-$ annihilation at the $Z^0$-peak. The
resulting diffractive structure function  compares very well with HERA data
(Fig.~\ref{fig:f2d3}). The GAL model also improves the description of
non-diffractive HERA data \cite{GAL-HERA}.

The GAL model can also be applied to $p\bar{p}$ to obtain diffractive $W$ 
and di-jet production through string rearrangements like in Fig.~\ref{fig:Wsci}.
The observed rates are reproduced quite well (Fig.~\ref{fig:Wsci}). 
However, the treatment of the `underlying event', which is a notorious 
problem in hadron-hadron scattering, introduces a larger uncertainty than 
for the SCI model \cite{SCI-W-jets}. 

\begin{figure}[tb]
\vspace*{-10mm}
\center{\epsfig{width=0.75 \columnwidth,file=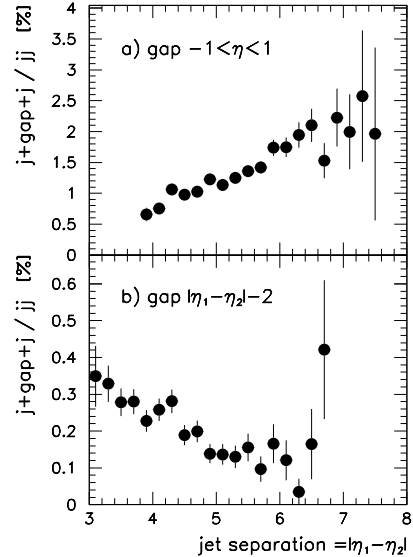}}
\vspace*{-12mm}
\caption{Relative rates of gaps between jets from GAL model simulations
of Tevatron $p\bar{p}$ with the gap size (a) fixed and 
(b) increasing with the jet separation \cite{SCI-W-jets}. 
}
\label{fig:gapsize}
\vspace*{-5mm}
\end{figure}

The Tevatron data on gaps between two high-$p_{\perp}$ jets are harder  to
understand. SCI does give such events, but at a too low rate.  The GAL model
can give the observed rate, but again with an uncertainty due to the treatment
of the underlying event.  The measured colour-singlet fraction in D0
\cite{D0-jgj} tends to increase with increasing jet separation,  whereas CDF
data \cite{CDF-jgj} shows no significant such effect.  However, the required
gap size is fixed to $-1<\eta < 1$.  Our Monte Carlo study shows an increase
with  jet separation with this fixed gap size, but a decrease when the gap
size  follows the jet separation (Fig.~\ref{fig:gapsize}). The proper
diffractive signature should rather be no suppression with increasing gap
size.  Thus, the exact gap definition is very important for the interpretation
and  this issue should therefore be examined further experimentally.

In conclusion, our  models for non-perturbative QCD dynamics in terms of 
varying colour string topologies give a satisfactory explanation of 
several phenomena, both diffractive and non-diffractive, 
thus providing a unified description of many different hadronic final states.

\end{document}